%% file: draft.tex
\def\editmode{0}

\def\bibfilenames{bibman_refs}

\def\singlenarrowcol{0}
\if\singlenarrowcol0   
    \documentclass[conference]{IEEEtran}

    \IEEEoverridecommandlockouts

\else
    \documentclass[onecolumn]{IEEEtran}
    \usepackage[pass]{geometry}
    \newgeometry{textwidth=252pt, textheight=672pt}
    
\fi

\usepackage{include_v7}

\usepackage{hyperref}

\usepackage{import}

\def\journal{0}

\usepackage{environ}

\if\journal0    
    \NewEnviron{journalonly}{%
    }
    \NewEnviron{conferenceonly}{%
    \BODY
    }
\else
    \if\journal1
        \NewEnviron{journalonly}{%
        \BODY
        }   
        \NewEnviron{conferenceonly}{%
        }  
    \else
        \if\journal2

        \fi
    \fi
\fi

\def\intermediatesteps{0}
\if\intermediatesteps1
    \NewEnviron{intermed}{%
        \color{purple}
        \BODY
    }

\else
    \NewEnviron{intermed}{%
    }

\fi

\input{notation}

\allowdisplaybreaks
\begin{document}

\title{ Bayesian Radio Map Estimation:\\Fundamentals and Implementation via \\Diffusion Models
    \thanks{This research has been funded in part by the Research Council of Norway under IKTPLUSS grant 311994.}
}

\author{
    \IEEEauthorblockN{Tien Ngoc Ha and Daniel Romero}

    \IEEEauthorblockA{\textit{
            Department of ICT},
        \textit{University of Agder}\\
        Grimstad, Norway \\
        \{tien.n.ha, daniel.romero\}@uia.no}
}


\maketitle

\begin{abstract}
    Radio map estimation (RME) is the problem of inferring the value of a
    certain metric (e.g. signal power) across an area of interest given a
    collection of measurements. While most works tackle this problem
    from a purely non-Bayesian perspective, some Bayesian estimators have been proposed. However, the latter focus on estimating the map itself -- the Bayesian standpoint is adopted mainly to exploit prior information or to capture uncertainty. This paper pursues a more general formulation, where the goal is to determine the posterior distribution of the map given the measurements. Besides handling uncertainty and allowing standard Bayesian estimates, solving this problem is seen to enable minimum mean square error estimation of arbitrary map functionals (e.g. capacity, bit error rate, or coverage area to name a few) while training only for power estimation. A general Bayesian estimator is proposed based on  conditional diffusion models and both the Bayesian and non-Bayesian paradigms are compared analytically and numerically to determine when the Bayesian approach is preferable.
\end{abstract}


\section{Introduction}
\label{sec:intro}

\begin{bullets}

    \blt[Motivation]
    \begin{bullets}
        \blt[overview RME] Radio map estimation (RME)~\cite{romero2022cartography} aims to construct spatial
        maps of radio frequency metrics — such as received signal strength,
        interference power, or channel gain — across a geographic region. Radio
        maps are typically estimated by spatially interpolating  measurements to
        predict the target  metric  at locations where no measurements were  collected. The most popular kind of radio maps are \emph{power maps}, which provide the power that a receiver would measure at each spatial location.







    \end{bullets}

    \blt[Literature]

    \begin{bullets}
        \blt[Non-Bayesian methods] Most works on  RME propose non-Bayesian estimators.
        \begin{bullets}
            \blt[Model-based]Traditional approaches include $k$-nearest neighbors, kernel ridge regression
            \cite{romero2017spectrummaps},
            sparsity-based inference~\cite{bazerque2010sparsity}, dictionary learning~\cite{kim2013dictionary},
            matrix completion
            \cite{schaufele2019tensor},
            and low-rank approximation~\cite{lee2014cartography} to name a few.
            \blt[Data-driven]More contemporary schemes are based on deep learning and include, for example, autoencoders~\cite{teganya2020rme,shrestha2022surveying}, U-Nets \cite{levie2019radiounet}, and transformers~\cite{viet2025spatial}.
        \end{bullets}






        \blt[Bayesian methods] Among Bayesian estimators, one can mention
        \begin{bullets}
            \blt[estimation]Kriging
            \cite{alayafeki2008cartography}, sparse Bayesian
            learning \cite{huang2014sparsebayesian, wang2024sparse}, deep learning~\cite{krijestorac2020deeplearning}, and schemes for Bayesian active learning
            \cite{eller2024uncertainty}.
            \blt[generation]Note that some recent works present Bayesian schemes for radio map \emph{generation} rather than estimation~\cite{luo2025denoising}.
        \end{bullets}
    \end{bullets}

    \blt[Focus]The reason why these schemes pursue a Bayesian approach  is twofold:
    \begin{bullets}%
        \blt[prior information](i) the Bayesian framework can accommodate various forms of side information (e.g. transmitter locations) via priors, and
        \blt[uncertainty](ii) Bayesian schemes can provide not only map estimates but also their associated uncertainty. This can be used e.g. to decide where to collect subsequent measurements.
    \end{bullets}
    However, existing works have not considered the estimation of \emph{map functionals}. Indeed, many relevant quantities in wireless communications depend on the spatial distribution of power and, therefore, they can be formalized as functionals that take a power map as input and return a scalar. For example, the coverage area is the area of the region in which the power map value exceeds a certain threshold. Another example would be the length of the trajectory of a robot or autonomous vehicle that avoids areas without connectivity. More examples include capacity, bit error rate, outage probability, and so on.

    \blt[contributions]
    \begin{bullets}
        \blt[formulation]The first contribution of this paper is a theoretical study of the \emph{general} Bayesian radio map estimation problem, where the goal is to estimate the posterior of the map rather than the map itself. This allows the estimation of arbitrary map functionals without retraining the estimator. Bayesian and non-Bayesian estimators are compared both analytically and numerically to determine when Bayesian estimators  are preferable.

        \blt[conditional diffusion model] The second contribution is a Bayesian
        estimator based on a conditional diffusion model. This estimator can sample from
        the posterior distribution of a radio map given the measurements. Relative to existing approaches, this scheme captures complex spatial dependencies without simplifying  assumptions such as Gaussianity.

    \end{bullets}

    \blt[paper structure]The rest of the paper is structured as follows. Sec.~\ref{sec:rme}  presents the model and reviews the non-Bayesian and Bayesian RME problem formulations. Sec.~\ref{sec:functional}  addresses the estimation of map functionals. Then, Secs.~\ref{sec:analytical} and \ref{sec:performance_evaluation}  compare Bayesian and non-Bayesian estimation theoretically and numerically, whereas Sec.~\ref{sec:diffusion} proposes the diffusion-based estimator. Finally,  Sec.~\ref{sec:conclusion} summarizes the main conclusions.

\end{bullets}


\section{Preliminaries}
\label{sec:rme}
\cmt{overview} This section introduces the mathematical model and reviews the  classical and Bayesian formulations of the RME problem.

\newcommand{\optional}[1]{}

\subsection{Model}
\label{sec:model}
\begin{bullets}
    \blt[model]
    \begin{bullets}
        \blt[region] Let $\area \subset \mathbb{R}^{\areadim}$ denote region of
        interest, where the map will be constructed. The dimension $\areadim$ is
        typically 2 or 3.
        \blt[transmitter]The  RF signals transmitted by an unknown number  of transmitters propagate through  $\area$ and contribute to
        \blt[map]the received power $\pow(\loc)$ at each
        location $\loc \in \area$. The function $\pow : \area \rightarrow
            \mathbb{R}$ that  assigns $\pow(\loc)$ to each $\loc$ is referred to as \emph{power map}, which is a special case of a radio map~\cite{romero2022cartography}. A power map depends on the transmitter locations, transmit power, their radiation patterns, the presence and physical nature of the objects in the environment, and other factors.

        \blt[channel-model]
        \optional{For example, when $\txnum=1$, the received power at a location $\loc$ is typically
            modeled as:
            \begin{align}
                \pow(\loc) = \ptx + \constgain - \spl(\loc) - \sshadow(\loc) - \sssf(\loc),
            \end{align}
            where $\ptx$ is the transmit power (in dBm), $\constgain$ captures
            antenna and system gain, $\spl(\loc)$ is the path loss, $\sshadow(\loc)$
            accounts for shadowing effects, and $\sssf(\loc)$ represents small-scale
            fading. \blt[assumptions] In practice, the small-scale fading component
            $\sssf(\loc)$ is often averaged out or considered negligible, enabling a
            focus on estimating the dominant, large-scale power structure.}

        \blt[receiver/measurements]To construct radio maps, measurements are collected at a set of
        locations $\left\{\loc_{\measind}\right\}_{\measind=1}^{\measnum}
            \subset \area$. The $\measind$-th measurement is modeled~as:
        \begin{align}
            \label{eq:measmodel}
            \powmeas{\measind} \define \pow(\loc_{\measind}) + \measnoise_{\measind},
        \end{align}
        where $\measnoise_{\measind}$ is additive
        zero-mean noise independent across $\measind$.
    \end{bullets}
\end{bullets}

\subsection{The Classical RME Problem Formulation}
\begin{bullets}

    \blt[overview]The most prevalent  formulation of  non-Bayesian RME  is as follows.
    \begin{bullets}%
        \blt[formulation]%
        \begin{bullets}%
            %
            \blt[Given]Given a set of noisy measurements
            $\measset\define \left\{(\loc_{\measind},
                \powmeas{\measind})\right\}_{\measind=1}^{\measnum}$,
            \blt[Find] the problem is to estimate $\pow(\loc)$ for all $\loc\in\area$.
        \end{bullets}
        Theoretically, the resulting estimate is a function that can be denoted as $\powest$. In practice, due to computational reasons, only the values of $\powest$ at a finite set of locations (e.g. a grid)  are typically obtained and stored.

        \blt[training dataset]For training, a dataset $\trainset\define \{
            \measset_1,\ldots,\measset_\realnum\}$ of measurement sets is available.
        Non-Bayesian radio map estimators essentially learn to interpolate measurements using
        such a training dataset.
    \end{bullets}
\end{bullets}

\label{sec:bayesian}



\subsection{The Bayesian RME Problem Formulation}
\label{sec:bayesianformulation}

\begin{bullets}
    \blt[overview]There are multiple ways of formulating the Bayesian RME problem. However, in all of them, the true radio map $\pow$ is random and its  probability distribution is typically unknown.
    \blt[formulation]The most general Bayesian formulation is the following:
    \begin{bullets}
        \blt[Given]given a  measurement set
        $\measset= \left\{(\loc_{\measind},
            \powmeas{\measind})\right\}_{\measind=1}^{\measnum}$,
        \blt[Find]obtain the posterior distribution of $\pow$.
    \end{bullets}
    \blt[posterior]This posterior distribution, denoted as $\pdf(\pow | \measset)$, is the mathematical representation of the \emph{joint} distribution of $\pow$ at an arbitrary set of locations given $\measset$. In other words, knowing $\pdf(\pow | \measset)$ is the same as knowing $\pdf(\pow(\evalloc_1),\ldots, \pow(\evalloc_\evallocnum) | \measset)$ for any $\evalloc_1,\ldots,\evalloc_\evallocnum$ and $\evallocnum$.

\end{bullets}







To understand the intuition behind this posterior, suppose for simplicity that
$\measset$ comprises a single noiseless measurement $(\loc_{1}, \powmeas{1})$
and let $\evallocnum=1$. The posterior $\pdf(\pow(\evalloc_1)|\measset)$ is essentially the probability density of  $\pow(\evalloc_1)$ for all possible maps that take the value $\powmeas{1}$ at $\loc_{1}$. In other words, if a dataset of realizations of $\pow$ is available, one could select those realizations whose value at $\loc_{1}$ is approximately $\powmeas{1}$. A histogram of the value that those maps take at $\evalloc_1$ will therefore converge to $\pdf(\pow(\evalloc_1)|\measset)$ as the number of realizations increases.

Following this intuition, it is easy to understand that, the larger $\measnum$,
the more concentrated $\pdf(\pow(\evalloc_1)|\measset)$ is around a value of
$\pow(\evalloc_1)$ since the selected realizations will be more similar. This is
precisely the notion of uncertainty in Bayesian~RME.

\subsection{Applicability of Bayesian RME}
Adopting the Bayesian formulation is motivated by the following reasons:

\begin{bullets}
    \blt[overview]

    \blt[application scenarios]
    \begin{itemize}
        \item\textbf{Bayesian point estimators.}
              In terms of the obtained information,  solving the  Bayesian
              problem is at least as good  as solving its non-Bayesian version since a solution to the former also allows point estimates. In particular,
              setting  $\evallocnum=1$ in such a solution yields $\pdf(\pow(\loc) | \measset)$.
              With this distribution, it is straightforward to obtain, for example, the
              \emph{minimum mean square error} (MMSE) estimator~\cite{kay1} of $\pow(\loc)$, which
              is nothing but the mean of this distribution, i.e., $\powest_{\text{MMSE}}(\loc) =
                  \expectation[\pow(\loc) | \measset]$. Another example is the \emph{maximum a
                  posteriori} (MAP) estimator, which is given by
              $\powest_{\text{MAP}}(\loc) = \arg\max_{\powdummy} \pdf(\pow(\loc)
                  = \powdummy | \measset)$.

        \item\textbf{Uncertainty.} Once an estimate like these is obtained,  one can use $\pdf(\pow(\loc) | \measset)$
              to quantify the uncertainty or error in this estimate. As explained in Sec.~\ref{sec:bayesianformulation}, this will be related to how spread this distribution is around the estimated values.

        \item\textbf{Prior information.} A Bayesian formulation facilitates capturing side information in the form of priors; see e.g. the references in Sec.~\ref{sec:intro}.

        \item\textbf{Map functionals.} While this has not been pointed out in
              the literature, the obtained $\pdf(\pow | \measset)$ can also be
              used for evaluating map functionals without retraining. In this
              case, $\evallocnum$ must be arbitrary. This is the subject of
              Sec.~\ref{sec:functional}.

    \end{itemize}

\end{bullets}

\section{Bayesian Map Functional Estimation}
\label{sec:functional}

\begin{bullets}
    \blt[Functionals] In many applications, one is interested in a quantity that depends on the power map but which is not directly power. For generality, this quantity will be represented by a functional $\metricfunc$ that takes a map $\pow$ as input and returns a real number.

    \subsection{Examples of Map Functionals}

    \blt[types]There are two broad classes of such functionals.
    \begin{bullets}%
        \blt[local]\emph{Local map functionals} depend on the map $\pow$ only through the value of $\pow$ at a single point, say $\evalloc$. In other words, they can be expressed as $\metricfunc(\pow)=\metricfun(\pow(\evalloc))$ for a real-valued function $\metricfun$.
        \begin{bullets}%
            \blt[Capacity]One example would be the \emph{capacity} of the channel between a transmitter and $\evalloc$. If noise is additive white Gaussian with variance $\noisepow$, it is given by
            \begin{align}
                \label{eq:capacity}
                \metricfunc(\pow) = \bandwidth \log_2\left[
                    1 + \frac{\pow(\evalloc)}{\noisepow}
                    \right],
            \end{align}
            where $\bandwidth$ denotes bandwidth.
            \blt[BER]Another example is the \emph{bit error rate} (BER). For example, for $\constelsize$-ary quadrature amplitude modulation (QAM), the BER is given by
            \begin{align}
                \label{eq:berqam}
                \metricfunc(\pow) = 3\left(1-\frac{1}{\sqrt{\constelsize}}\right)\erfc\left(\sqrt{\frac{3\pow(\evalloc)}{(\constelsize-1)\noisepow}}\right),
            \end{align}
            where $\erfc$ is the Gaussian error function.


            \blt[SINR]For another example, suppose that  the received signal power is fixed to a certain value $\signalpow$  (e.g. via an automatic gain control loop involving the transmitter) and $\pow(\loc)$ quantifies interference at $\loc$, then the \emph{signal to interference plus noise ratio} (SINR) at $\evalloc$ is
            \begin{align}
                \metricfunc(\pow) = \frac{\signalpow}{\pow(\evalloc)+\noisepow}.
            \end{align}
            \blt[Outage]The last example here of a local map functional is the \emph{outage indicator} $\metricfunc(\pow) = \indicator[{\pow(\evalloc) < \threshold}]$, where  $\threshold$ is a threshold and $\indicator[\cdot]$ is 1 if the condition inside brackets holds and 0 otherwise.

        \end{bullets}

        \blt[global]On the other hand, \emph{global map functionals} are those that are not local. They may depend on the entire $\pow$.
        \begin{bullets}
            \blt[coverage area]For example, the \emph{coverage area} is defined as the
            total area where the received power exceeds a certain threshold
            $\threshold$, i.e.,
            \begin{align}
                \label{eq:service_area_nonbayesian}
                \metricfunc(\pow) = \int_{\area} \indicator[{
                            \pow(\loc) \geq \threshold}] d\loc,
            \end{align}
            \blt[Arrival time]Another example arises by considering a robot or autonomous vehicle that plans a path
            through points $\loc\in \area$ satisfying $\pow(\loc)\geq \threshold$. The
            trajectory is, therefore, determined by $\pow$ and, as a result,  one can set $\metricfunc(\pow)$ to be
            the
            time it takes to traverse~it.
        \end{bullets}
    \end{bullets}

    \subsection{Map Functional Estimation}
    \label{sec:mapfunctionalestimation}
    \blt[metric estimation]Having introduced map functionals, the next step is
    to explore how they can be estimated.
    \begin{bullets}
        \blt[problem] In particular, the problem is
        \begin{bullets}%
            \blt[requested]to estimate $\trueme \define \metricfunc(\pow)$
            \blt[given]given $\metricfunc$ and $\measset$.
        \end{bullets}

        \blt[approaches]
        \begin{bullets}
            \blt[Non-bayesian estimate: f(power estimate)]
            \begin{bullets}
                \blt[description]The most immediate approach that  one can think of is to use any estimator of $\pow$ to obtain  an estimate $\powest$ and then estimate $\trueme$ as
                $\metricfunc(\powest)$.
                \blt[limitation \ra MMSE not minimal ]Unfortunately, this
                approach is suboptimal. To see why, note that conventional radio
                map estimators are typically trained by minimizing a square
                loss; see e.g. the references in Sec.~\ref{sec:intro}. From a
                Bayesian perspective, a square loss approximates the mean square
                error (MSE) and, as a result, these schemes approximately yield
                MMSE estimates if sufficiently trained. With $\powest(\loc)$
                denoting one of these estimates at $\loc$, it follows~\cite{kay1} that
                $\powest(\loc)\approx \expectation[\pow(\loc)|\measset]$. The
                aforementioned estimate of $\trueme $
                can therefore be expressed as  $ \metricfunc(\powest)\approx
                    \nonbayme \define\metricfunc(\expectation[\pow|\measset])$,
                where $\expectation[\pow|\measset]$ denotes the function $\loc
                    \mapsto \expectation[\pow(\loc)|\measset]$.
                \blt[MMSE estimator of f]In turn, the actual MMSE estimator of
                $\trueme$ is
                $\bayme\define\expectation[\metricfunc(\pow)|\measset]\neq \nonbayme$, which shows that $\nonbayme$ does not minimize the MMSE.

                \blt[terminology] To facilitate the explanation, $\nonbayme$ and
                $\bayme$ will be respectively referred to as the \emph{non-Bayesian}
                and \emph{Bayesian estimates}.

                \blt[OK if linear]Observe that, if $\metricfunc$ is  linear,
                then $\nonbayme=\bayme$. However, for nonlinear $\metricfunc$,
                these  estimates may significantly differ.
                \blt[example if f convex/concave]For example,  if $\metricfunc$ is a local map functional associated with a strictly convex $\metricfun$, it follows
                by
                Jensen's inequality that
                \begin{align}
                    \metricfun(\expectation[\pow(\loc)|\measset]) < \expectation[\metricfun(\pow(\loc))|\measset],
                \end{align}
                which implies that $\nonbayme < \bayme$. Conversely, if
                $\metricfun$ is strictly concave, then $\nonbayme > \bayme$.

            \end{bullets}

            \subsection{Obtaining Bayesian Map Functional Estimates}
            \label{sec:obtainingbayesianmfe}
            \blt[train from f(dataset)]
            \begin{bullets}
                \blt[description]In the case of local functionals, it is interesting to see that  $\bayme$ can be obtained without solving the Bayesian RME problem: it suffices to apply  a conventional radio map estimator trained on a dataset where each measurement $\powmeas{\measind}$ is replaced with $\metricfun(\powmeas{\measind})$. In this way, instead of power maps, the algorithm learns to estimate capacity maps, BER maps, coverage maps, etc.
                \blt[limitation: train/have a separate estimator for each f]The limitations of such an approach are that (i) a systematic error is introduced when $\metricfun$ is not linear  because $\metricfun(\powmeas{\measind}) = \metricfun(\pow(\loc_\measind) + \measnoise_{\measind})$ and, as a result, the effective noise is  not unbiased; and (ii) one needs to separately train for each $\metricfun$ of interest.
            \end{bullets}

            \blt[Bayesian estimate]For these reasons, actually solving the Bayesian RME problem may be more convenient. Once this is done, $\bayme$ may be obtained in two ways:
            \begin{itemize}
                \item\textbf{Integral form.}
                      \begin{bullets}
                          \blt[expression]Some schemes  provide $\pdf(\pow(\loc)|\measset)$ in closed form, e.g. as a Gaussian distribution~\cite{krijestorac2020deeplearning}. Hence,  if $\metricfunc$ is a local map functional, $\bayme$ can be computed~as
                          \begin{align}
                              \bayme = \int \metricfun( \powdummy) \pdf\Big(\pow(\evalloc)=\powdummy \Big| \measset\Big) d\powdummy.
                          \end{align}

                      \end{bullets}

                \item\textbf{Sample form.}
                      \begin{bullets}
                          \blt[limitation of integral: posterior often not available in closed form]Unfortunately, the integral above may be challenging to evaluate in closed form and $\pdf(\pow(\loc)|\measset)$ may not even be available. In fact, many Bayesian methods in machine learning will represent a distribution $\pdf(\pow|\measset)$ \emph{implicitly} by providing a means of \emph{sampling} from this distribution rather than a closed-form expression.
                          \blt[description]In these cases, one approximates $\bayme$ as
                          \begin{align}
                              \label{eq:bayesianfuncavg}
                              \bayme \approx \frac{1}{\sampnum}\sum_{\sampind=1}^{\sampnum} \metricfunc(\pow_\sampind),
                          \end{align}
                          where $\pow_1,\ldots,\pow_\sampind$ are samples from $\pdf(\pow|\measset)$.
                      \end{bullets}

            \end{itemize}

        \end{bullets}

    \end{bullets}

\end{bullets}

\section{Analytical Comparison of Bayesian and non-Bayesian Estimators}
\label{sec:analytical}
\begin{bullets}
    \blt[overview]While the non-Bayesian RME problem has already been studied theoretically~\cite{romero2024theoretical}, a fundamental analysis of the Bayesian RME problem has never been undertaken.
    To contribute in this direction, this section provides an example of application scenario where Bayesian estimators are analytically shown to perform strictly better than non-Bayesian estimators.

    \subsection{Estimation Setup}

    \blt[goal]For simplicity, the problem is the 1D counterpart of the problem of estimating coverage area. Specifically, the region of interest is the x-axis, i.e., the set of points of the form $\loc = [\locx; 0; 0]$, $\locx\in \rfield$. There, one is interested in the \emph{coverage length}, which is defined along the lines of \eqref{eq:service_area_nonbayesian} for a  threshold $\threshold> 0$ as $\metricfunc(\pow) = \int_{-\infty}^{\infty} \indicator[{
                    \pow([\locx; 0; 0]) \geq \threshold}] d\locx$.

    \blt[map distribution]It remains to specify a distribution over $\pow$. To this end, a single transmitter operating in free space will be considered, which allows tractability.
    \begin{bullets}
        \blt[Friis]If $\txloc = [\txlocx;
            \txlocy; \txlocz]$ denotes the  transmitter  location, the power received at  $\loc$ is given by
        \begin{align}
            \pow(\loc)\define \pow(\loc;\txloc) \define \frac{\txpow \txantgain \rxantgain \wavelen^2}{(4\pi)^2 \|\loc - \txloc\|^2} = \frac{\txpowwave}{\|\loc - \txloc\|^2},
            \label{friis}
        \end{align}
        where
        \begin{bullets}
            \blt[alpha]$\txpowwave \define \txpow
                \txantgain\rxantgain (\wavelen/4\pi)^2$ is a constant defined in terms of
            \blt[tx pow] the transmitted power $\txpow$,
            \blt[ant gains] the transmitter
            and receiver antenna gains $\txantgain$ and $\rxantgain$,
            \blt[wavelength] and the wavelength $\wavelen$.
        \end{bullets}
        \blt[1D scenario]For convenience,  the notation in \eqref{friis} will be specialized to points of the form $\loc = [\locx; 0; 0]$ as
        \begin{align}
            \pow(\locx; \txlocx) = \frac{\txpowwave}{(\locx - \txlocx)^2 + \proj^2},
            \label{friis_simplified_1d}
        \end{align}
        where $\proj = \sqrt{\txlocy^2 + \txlocz^2}$ is the distance from the
        transmitter to the x-axis.
        \blt[distribution]A distribution over $\pow$ will  then be defined by specifying that $\proj$ is fixed and $\locx$ uniformly distributed over a sufficiently large interval.
    \end{bullets}

    \blt[True coverage length]The true coverage length can be readily computed by finding the length of the interval $\{\locx~:~\pow(\locx; \txlocx)\geq \threshold\}$.
    \begin{bullets}
        \blt[Service region]When $ \threshold  < \frac{\txpowwave}{\proj^2}
        $, a point $\locx$ can be seen to be in this set if and only if
        \begin{bullets}
            \blt[greater or equal than 0]
            \begin{align}
                \txlocx - \sqrt{\frac{\txpowwave}{\threshold}\left(\frac{\threshold}{\txpowwave}-\proj^2\right)} \leq \locx
                \leq \txlocx + \sqrt{\frac{\txpowwave}{\threshold}\left(\frac{\threshold}{\txpowwave}-\proj^2\right)}.
            \end{align}
            \blt[less than 0]Otherwise, the set is empty.
        \end{bullets}
        \blt[True coverage length]Thus,  the  true coverage length~is
        \begin{align}
            \label{eq:service_length_true}
            \trueme = \metricfunc(\pow) =\begin{cases}
                                             2\sqrt{\frac{\txpowwave}{\threshold}\left(\frac{\threshold}{\txpowwave}-\proj^2\right)}
                                               & \text{if }\threshold  < \frac{\txpowwave}{\proj^2} \\
                                             0 & \text{otherwise}.
                                         \end{cases}
        \end{align}
    \end{bullets}

    \begin{figure}[t]
        \centering
        \includegraphics[width=.9\linewidth]{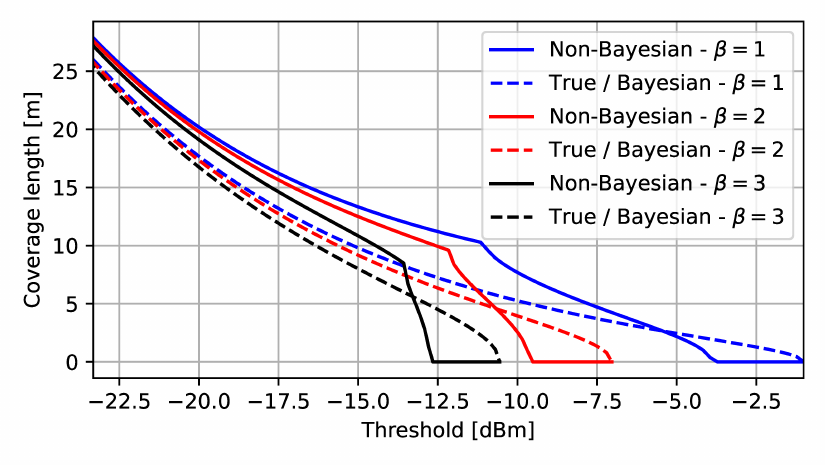}
        \caption{Non-Bayesian estimates compared with the true coverage length for the scenario of Sec.~\ref{sec:analytical}
            ($\txpow=30$ dBm, $\txantgain = \rxantgain = 0$ dBi, $\txlocx= 3$,  2.4 GHz carrier frequency).}
        \label{fig:analyticalcomparison}
    \end{figure}

    \subsection{The Bayesian and non-Bayesian estimates}
    \blt[Bayesian]Recall from Sec.~\ref{sec:mapfunctionalestimation} that the Bayesian estimate is given by $\bayme\define\expectation[\metricfunc(\pow)|\measset]$. Noting from \eqref{eq:service_length_true} that $\metricfunc(\pow)$ does not depend on any random quantity establishes that $\bayme\define\expectation[\metricfunc(\pow)|\measset]=\metricfunc(\pow) =\trueme$. In words, this means that \emph{the Bayesian estimator produces the exact coverage length}, i.e.,  it has no error in this setup. And this holds regardless of $\measnum$.  This was expected because, with sufficient training, a Bayesian estimator will generate samples that look like the true maps. What changes from sample to sample is their position on the x-axis, but their coverage length  is  the~same.

    \blt[Non-Bayesian]
    \begin{myproposition}
        \emph{
            \begin{bullets}%
                \blt[Measurements]Suppose that $\measset=\{([\locx_1;0;0], \powmeas{1})\}$, where $\powmeas{1}$ has no noise,  and
                \blt[Definitions]let $A\define-{\threshold}/{\txpowwave}$,
                $B\define1+{2\threshold}/{\powmeas{1}} -
                    {4\threshold\proj^2}/{\txpowwave}$,  $C\define{\txpowwave}/{\powmeas{1}} - {\threshold\txpowwave}/{\powmeas{1}^2}$, $D \define B^2 - 4AC$,  and $\delt\define\sqrt{\txpowwave/\powmeas{1}-\proj^2}$.
                \blt[Non-B coverage length]The non-Bayesian estimate $\nonbayme
                    \define\metricfunc(\expectation[\pow|\measset])$ of the coverage length is
                \begin{align}
                    \label{eq:service_length_analytical_nonb}
                    \nonbayme =
                    \begin{cases}
                        2 \sqrt{\frac{\txpowwave}{\threshold}\left(\frac{B + \sqrt{D}}{2}\right)}       & \text{if }\threshold < \powmeas{1}                                                \\
                        2\bigg[ \sqrt{\frac{\txpowwave}{\threshold}\left(\frac{B + \sqrt{D}}{2}\right)} & -~ \sqrt{\frac{\txpowwave}{\threshold}\left(\frac{B - \sqrt{D}}{2}\right)} \bigg]
                        \\& \text{if }\powmeas{1} \leq \threshold < \powestmax \\
                        0                                                                               & \text{if }\powestmax \leq \threshold ,
                    \end{cases}
                \end{align}
                where $\powestmax\define \argmax_\locx (\pow(\locx; -\delt) + \pow(\locx; \delt) )/2$.
            \end{bullets}%
        }
    \end{myproposition}
    \begin{IEEEproof}
        The proof is omitted due to space constraints.
    \end{IEEEproof}


    \blt[Interpretation]
    \begin{bullets}
        \blt[Expressions]Observe that \eqref{eq:service_length_true} and
        \eqref{eq:service_length_analytical_nonb} do not coincide and, therefore,
        the non-Bayesian estimator incurs estimation error.
        \blt[figure]Since assessing the magnitude of this error by comparing
        \eqref{eq:service_length_true} and \eqref{eq:service_length_analytical_nonb}
        is not straightforward, Fig.~\ref{fig:analyticalcomparison} plots the
        non-Bayesian  estimate along with the true coverage length, which coincides
        with the Bayesian estimate. For the selected  parameter values, the error is sometimes over 50\%.

    \end{bullets}

    \blt[more realistic models]Finally, it is worth emphasizing that the example
    presented in this section was chosen intentionally simple to guarantee
    tractability. A comparison with more realistic models requires  numerical means, and this is the subject of
    Sec.~\ref{sec:performance_evaluation}.
\end{bullets}

\section{Diffusion models for Bayesian RME}
\label{sec:diffusion}
\begin{figure}[t]
    \centering
    \includegraphics[width=\linewidth]{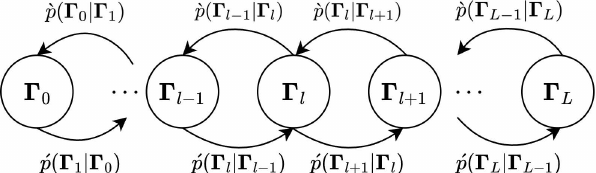}
    \caption{The forward and reverse processes of a diffusion model.}
    \label{fig:vdm}
\end{figure}

\begin{bullets}
    \blt[overview] This section proposes an estimator for the Bayesian RME problem. It is based on \emph{diffusion models}, a class of generative models for learning complex data distributions that is inspired by  diffusion
    processes~\cite{ho2020diffusion}. They have shown remarkable success in various domains, including
    image generation and audio synthesis~\cite{luo2022diffusion}.

    \blt[Unconditional Diffusion models]
    \begin{bullets}
        \blt[overview]To understand diffusion models, it is convenient to first consider \emph{unconditional} diffusion models. Once  trained over a set of vectors, these models can generate further samples with the same distribution. For example, if trained on pictures of cats, they will generate more pictures of cats.
        \blt[initial]To see how this is done, let  $\vx_0$ denote one of the elements of the training dataset. In the case of RME, $\vx_0$ would be a matrix obtained by evaluating a radio map $\pow$ at a grid of spatial locations.
        %
        \blt[forward] Using $\vx_0$, a sequence  $\vx_1, \vx_2,
            \ldots, \vx_L$ is generated according to the entrywise uncorrelated Gaussian distribution  $\acute{q}(\vx_\stepind|\vx_{\stepind-1}) =
            \mathcal{N}(\vx_\stepind;
            \sqrt{\alpha_\stepind}
            \vx_{\stepind-1}, 1-\alpha_\stepind)$, where $\alpha_\stepind \in
            [0, 1]$ is the variance of each entry at step $\stepind$. This \emph{forward process} is illustrated in Fig.~\ref{fig:vdm}. For large $\stepind$, $\vx_\stepind$ becomes indistinguishable from Gaussian noise.

        %
        %
        \blt[reverse] Note that a sequence like this could be generated for every $\vx_0$ in the dataset.
        \blt[training] Training a diffusion model means training a  neural network
        to  predicts  $\hat{\vx}_{\stepind}$ given $\stepind$ and
        $\vx_{\stepind+1}$.  To this end, a square loss can be used
        \cite{luo2022diffusion}. Once this network is trained, one can generate a sample of $\vx_{L}$ using a Gaussian distribution and then recursively applying the neural network to obtain a sample of $\vx_0$.

    \end{bullets}

    \blt[Conditional Diffusion Model for RME]
    \begin{bullets}
        \blt[overview]While useful to grasp the intuition, unconditional diffusion models are not suitable for RME because they can just be used to generate samples of the \emph{prior} $\pdf(\pow)$. If one wishes to generate samples from the posterior $\pdf(\pow|\measset)$, \emph{conditional} diffusion models should be used instead. In models of this kind, the forward process is identical to that in unconditional diffusion models. However, in the reverse process, $\measset$ is also passed to the neural network. Intuitively, this allows the model to produce  samples that are consistent with the measurements.
        In practice, besides $\measset$, the network can also be given a building mask if available along the lines of~\cite{teganya2020rme}.

    \end{bullets}
\end{bullets}

\section{Numerical Experiments}
\label{sec:performance_evaluation}

\cmt{overview}
This section presents numerical experiments that (i) compare Bayesian and
non-Bayesian estimators, and (ii) assess the performance of the
diffusion-based scheme from Sec.~\ref{sec:diffusion}.


\subsection{Experimental Setup}
\label{sec:experiments}

\begin{bullets}
    \blt[Scenario Configurations]
    \begin{bullets}
        \blt[overview] Two distinct scenarios are considered: \emph{line of sight} (LoS) and \emph{non-line of sight}
        (NLoS).
        \blt[common]The following parameters are common to both.
        \begin{bullets}%
            \blt[area]The region $\area$ is a
            $160\,\text{m} \times 96\,\text{m}$ rectangle.
            \blt[tx power] The transmit power is $44\,\text{dBm}$,
            \blt[gains] the antenna gains $0\,\text{dBi}$,
            \blt[freq.] and the carrier  frequency $3.5\,\text{GHz}$.
            \blt[Tx height] The transmitter locations are uniformly distributed on a horizontal plane of
            height  $10\,\text{m}$.
            \blt[Rx height] The measurements are collected on a horizontal plane of height  $1\,\text{m}$. The ground is at $0$ m height.
        \end{bullets}

        \blt[LoS Scenario]In the LoS scenario, maps are defined over a  $32
            \times 32$ rectangular grid and propagation adheres to \eqref{friis}.
        \blt[NLoS Scenario]In the NLoS scenario, maps are defined over a  $64
            \times 64$ grid. Two transmitters are deployed and propagation is
        simulated via the Sionna library~\cite{sionna2021}, which accounts
        for multipath propagation and environmental obstructions such as
        buildings.
    \end{bullets}

    \blt[Estimators]Three estimators will be considered. All of them are trained
    to \emph{estimate power} on a dataset with $\realnum=100,000$ maps for the LoS scenario (40,000 of them  for validation) and $\realnum=60,000$ for the NLoS scenario (24,000 of them for validation).
    \begin{bullets}
        \blt[Diffusion]The first estimator is the diffusion-based scheme from
        Sec.~\ref{sec:diffusion} with
        \begin{bullets}%
            \blt[No. time steps] $\numsteps = 1,000$
            steps,
            \blt[decay fator]a noise schedule where  the noise coefficient
            $\alpha_\stepind$ decays linearly from $\alpha_1 = 0.9999$ to
            $\alpha_\numsteps = 0.98$,
            \blt[architecture] and the  U-Net network architecture in
            \cite{ho2020diffusion} without attention layers. The
            number of  parameters of this network is
            approximately $2.7$ million.
        \end{bullets}%
        \blt[UnetStd]The second estimator is the one from
        \cite{krijestorac2020deeplearning}, where
        $\pdf(\pow(\loc_1),\ldots,\pow(\loc_\evallocind),\ldots,\pow(\loc_\evallocnum)|\measset)$
        is a Gaussian distribution uncorrelated along $\evallocind$ whose mean
        and diagonal covariance are obtained by evaluating a convolutional U-Net
        with approximately $5.5$
        million parameters on $\measset$.
        %
        \blt[Kriging]The third estimator is simple Kriging with the Gudmundson
        shadowing model \cite{shrestha2022surveying}. The parameters
        of the estimator are  the two parameters of this model.

    \end{bullets}

\end{bullets}

\subsection{Experiments}

\begin{bullets}
    \blt[Sampling]Fig. \ref{fig:NLoS_compare} shows samples from the posterior
    inferred with the three estimators from $\measnum=5$ measurements
    in the NLoS scenario.
    \begin{bullets}%
        \blt[diffusion]The samples generated using the diffusion model look like
        actual radio maps. The shadowing patterns are highly realistic. The differences
        between samples are due to the uncertainty that remains after observing
        the map at the measurement locations. For larger $\measnum$, these
        differences vanish and the samples would gradually become more similar
        to the true map.
        \blt[krijestorac]With Krijestorac et al., the samples
        are noisy because of the assumption that the map values are spatially
        uncorrelated.
        \blt[kriging]With Kriging, the samples are overly smooth. As a classical
        interpolation scheme, it requires a significantly larger number of
        measurements to accurately estimate  maps.
    \end{bullets}

    \begin{figure}[t]
        \centering
        \includegraphics[width=\linewidth]{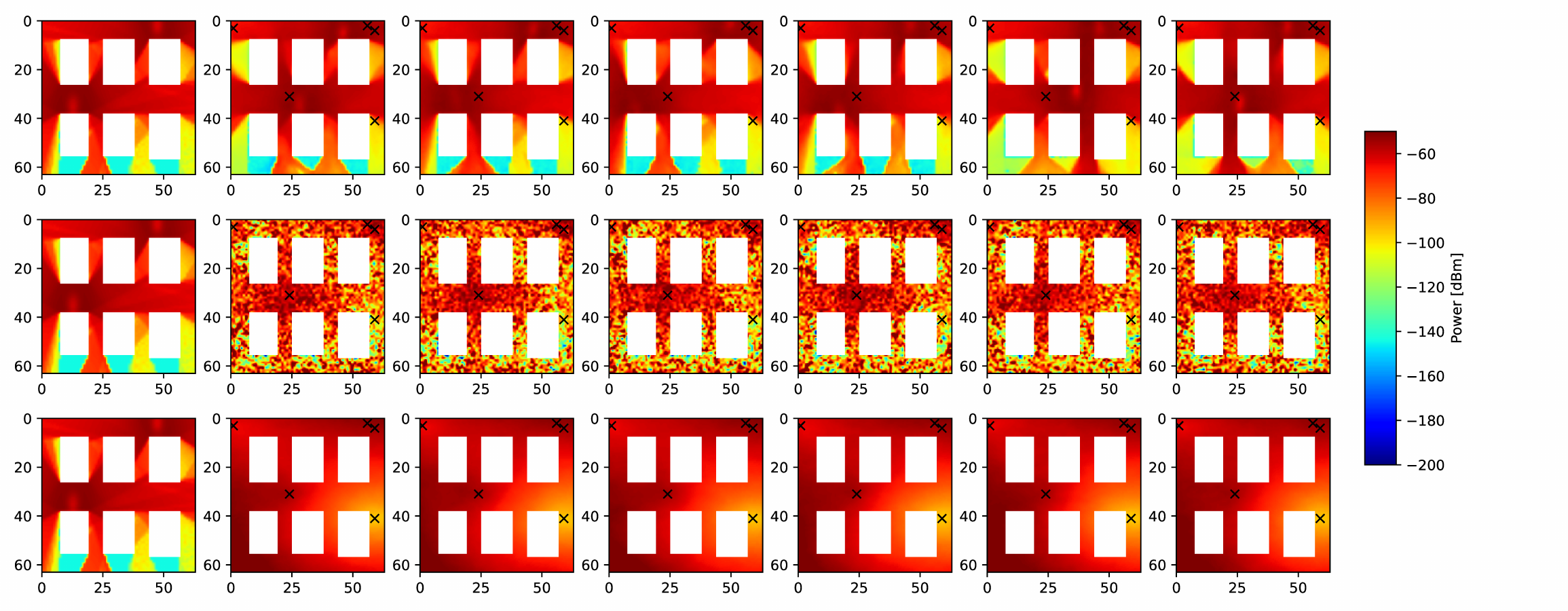}
        \caption{Samples given $5$ measurements in the  NLoS scenario (top:
            Diffusion, middle: Krijestorac et al., bottom: Kriging). Black crosses are measurement locations and white rectangles are buildings seen from above.}
        \label{fig:NLoS_compare}
    \end{figure}

    \blt[Bayesian vs. non-Bayesian]Fig.~\ref{fig:LoS_capacity} compares $\bayme$
    and $\nonbayme$ when $\metricfunc$ is given by \eqref{eq:capacity} and
    \eqref{eq:berqam}, where $\noisepow=-30$ dBm and $\constelsize=256$.
    In both cases, $10,000$ samples are obtained using the diffusion
    model. While $\bayme$ is directly implemented as \eqref{eq:bayesianfuncavg},
    obtaining $\nonbayme \define\metricfunc(\expectation[\pow|\measset])$
    requires $\expectation[\pow|\measset]$, which is computed by averaging the
    samples. As anticipated, the Bayesian estimator performs better. In the case
    of capacity, the error may differ by roughly 1 Mbps. The improvement in the
    case of the BER is more subtle.

    \begin{figure}[t]
        \centering
        \includegraphics[width=\linewidth]{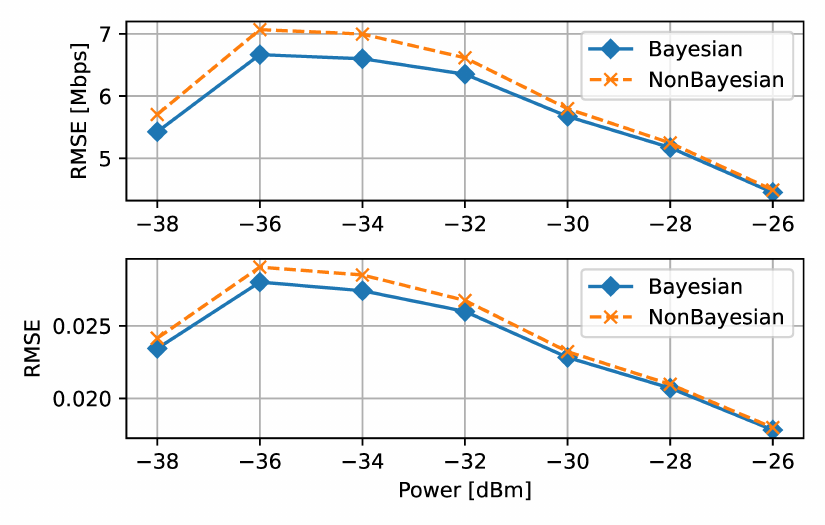}
        \caption{Bayesian vs. on Bayesian estimation of capacity (top) and BER (bottom) in the LoS scenario with one transmitter.}
        \label{fig:LoS_capacity}
    \end{figure}

    \blt[PAE]The last two experiments assess how well the considered schemes
    estimate coverage area, which is defined by $\metricfunc$ in \eqref{eq:service_area_nonbayesian}.
    \begin{bullets}%
        \blt[Percentage of AE] The performance metric is the \emph{percentage of area error} (PAE), which for an estimate $\me$ is defined as $\text{PAE} = 100 |\truemetric-\me|/\truemetric$.
        \blt[coverage estimator]As an additional benchmark, a coverage estimator   as described at the beginning of Sec.~\ref{sec:obtainingbayesianmfe} is included.
        Each training measurement $\powmeas{\measind}$ is replaced with $
            \indicator[{ \powmeas{\measind} \geq \threshold}]$ and a classification
        loss is used. The algorithm is implemented using a U-Net with the same
        architecture as the second estimator.

        \blt[Figures]Figs. \ref{fig:LoS_coverage_2Tx} and
        \ref{fig:NLoS_coverage} respectively show the PAE in the LoS and NLoS scenarios.
        \begin{bullets}%
            \blt[diffusion]As expected, the Bayesian diffusion estimator offers the
            best performance. Its non-Bayesian counterpart performs similarly for a
            sufficiently large number of measurements. This is because, as $\measnum$
            increases, uncertainty decreases and the samples become more similar to each
            other.
            \blt[coverage estimator]The coverage estimator, although specifically
            trained for this setup, yields worse performance because it relies on
            binary data.
            \blt[Krijestorac]The scheme by Krijestorac et al. seems to perform
            better when the Bayesian estimate is used. Note that, in this case, the
            estimate depends on both the mean and variance returned by the network.
            In contrast, the non-Bayesian estimator depends  solely on
            the mean, and this is seen to be detrimental except in the NLoS scenario
            when the number of measurements is large. This is expected since this
            model assumes that the posterior is uncorrelated Gaussian, which is a
            significant deviation from the actual distribution; cf.
            Fig.~\ref{fig:NLoS_compare}.

            \blt[Kriging]
        \end{bullets}

    \end{bullets}

    \begin{figure}[t]
        \centering
        \includegraphics[width=\linewidth]{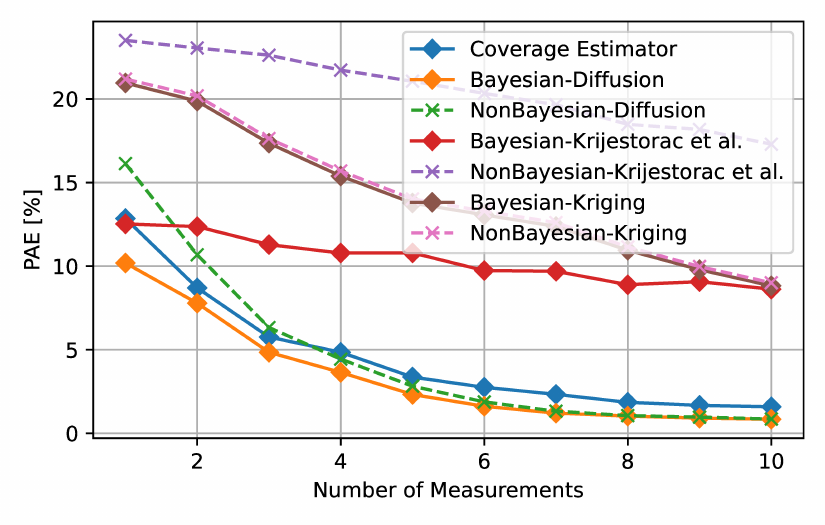}
        \caption{PAE of the estimators in the LoS scenario with two transmitters ($\threshold
                =-35~\text{dBm}$).}
        \label{fig:LoS_coverage_2Tx}
    \end{figure}

    \begin{figure}[t]
        \centering
        \includegraphics[width=\linewidth]{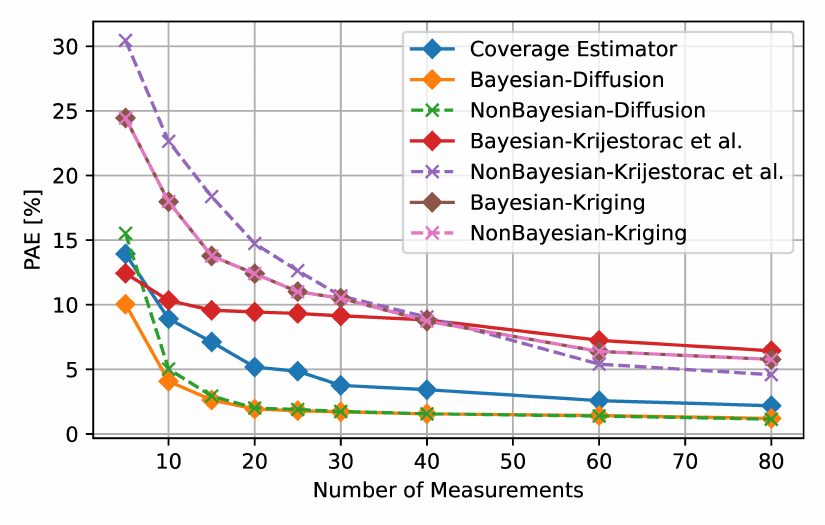}
        \caption{PAE of the estimators in the NLoS scenario ($\threshold=-95~\text{dBm}$).}
        \label{fig:NLoS_coverage}
    \end{figure}

\end{bullets}

\section{Conclusion}
\label{sec:conclusion}

This paper offered a new perspective of the Bayesian RME problem where the
posterior is inferred from measurements to estimate map functionals. Bayesian
and non-Bayesian estimators were extensively discussed and compared, both
analytically and numerically. Remarkably, Bayesian RME estimators can be used to obtain MMSE estimates of arbitrary map functionals, whereas their non-Bayesian counterparts can only produce MMSE estimates for linear map functionals. A novel Bayesian estimator was proposed using diffusion models and seen to outperform non-Bayesian estimators in numerical experiments.



\printmybibliography

\end{document}

%% file: notation.tex
\newcommand{\area}{\hc{\mathcal{X}}} 
\newcommand{\areadim}{\hc{d}} 
\newcommand{\txnum}{\hc{S}} %
\newcommand{\ptx}{\hc{P_{\text{Tx}}}} 
\newcommand{\spl}{\hc{s_{\text{PL}}}} 
\newcommand{\sshadow}{\hc{s_{\text{S}}}} 
\newcommand{\sssf}{\hc{s_{\text{SSF}}}} 
\newcommand{\loc}{{\hc{{\bm{r}}}}} 
\newcommand{\evalloc}{{\hc{{\bbm{r}}}}} 
\newcommand{\constgain}{{\hc{G}}} 
\newcommand{\pow}{{\hc{\gamma}}} 
\newcommand{\powdummy}{{\hc{\check\gamma}}} %

\newcommand{\measind}{{\hc{n}}} 
\newcommand{\measnum}{\hc{N}} 
\newcommand{\measnoise}{\hc{\zeta}} 
\newcommand{\measset}{\hc{\mathcal{M}}} 

\newcommand{\trainset}{\hc{\mathcal{D}}} 
\newcommand{\realnum}{{\hc{M}}} %

\newcommand{\constelsize}{{\hc{M}_\text{c}}}

\newcommand{\evallocnum}{{\hc{K}}}
\newcommand{\evallocind}{{\hc{k}}}

\newcommand{\powmeas}[1]{\hc{\tilde{\pow}}_{#1}} 
\newcommand{\powest}{{\hc{\hat \pow}}} 
\newcommand{\powestmax}{{\hc{\hat \pow^*}}} %
\newcommand{\expectation}{{\hc{\mathbb{E}}}} 

\newcommand{\metricfunc}{\hc{F}} 
\newcommand{\metricfun}{\hc{f}} 
\newcommand{\sampnum}{{\hc{J}}} 
\newcommand{\sampind}{{\hc{j}}} 
\newcommand{\pdf}{\hc{p}} 
\newcommand{\threshold}{\hc{\zeta }} 
\newcommand{\trueme}{\hc{\phi}} 
\newcommand{\truemetric}{\trueme} 
\newcommand{\nonbayme}{\hc{\hat\phi_{\text{NB}}}} 
\newcommand{\bayme}{\hc{\hat\phi_{\text{B}}}} 
\newcommand{\me}{\hc{\hat\phi}} 

\newcommand{\indicator}{\hc{\mathbb{I}}} 

\newcommand{\bandwidth}{\hc{B}} %
\newcommand{\erfc}{\hc{\text{erfc}}} 
\newcommand{\noisepow}{\hc{\sigma^2}} %
\newcommand{\signalpow}{\hc{\sigma^2_\text{s}}} %

\newcommand{\txloc}{{\hc{{\acute{\bm{r}}}}}} 
\newcommand{\txpow}{{\hc{P_{\text{Tx}}}}} 
\newcommand{\txantgain}{{\hc{G_{\text{Tx}}}}} 
\newcommand{\rxantgain}{{\hc{G_{\text{Rx}}}}} 
\newcommand{\wavelen}{{\hc{\lambda}}} 
\newcommand{\txpowwave}{{\hc{\alpha}}} 
\newcommand{\locx}{{\hc{x}}} 
\newcommand{\txlocx}{{\hc{{\acute{x}}}}} 
\newcommand{\txlocy}{{\hc{{\acute{y}}}}} 
\newcommand{\txlocz}{{\hc{{\acute{z}}}}} 
\newcommand{\proj}{{\hc{\beta}}} 
\newcommand{\delt}{{\hc{\Delta}}} 

\newcommand{\vx}{{\hc{\bm \Gamma }}}

\newcommand{\stepind}{{\hc{l}}}
\newcommand{\numsteps}{{\hc{L}}}